\documentclass[prb,amsmath,twocolumn,superscriptaddress,showpacs]{revtex4}
\usepackage{epsfig}
\usepackage{amssymb}
\usepackage{amsmath}
\usepackage{amsfonts}
\usepackage[T1]{fontenc}
\usepackage{bm}
\usepackage{graphicx}
\usepackage{color}

\begin{document}
\title{Quantum and Classical in Adiabatic Computation}

\author{P.~J.~D. Crowley}
\affiliation{London Centre for Nanotechnology, University College London, Gordon St., London, WC1H 0AH, United Kingdom}

\author{T.   \DJ uri\' c}
\affiliation{London Centre for Nanotechnology, University College London, Gordon St., London, WC1H 0AH, United Kingdom}

\author{W. Vinci}
\affiliation{London Centre for Nanotechnology, University College London, Gordon St., London, WC1H 0AH, United Kingdom}

\author{P.~A. Warburton}
\affiliation{London Centre for Nanotechnology, University College London, Gordon St., London, WC1H 0AH, United Kingdom}

\author{A.~G. Green}
\affiliation{London Centre for Nanotechnology, University College London, Gordon St., London, WC1H 0AH, United Kingdom}

\date{\today}

\begin{abstract}
Adiabatic transport provides a powerful way to manipulate quantum states. By preparing a system in a readily initialised state and then slowly changing its Hamiltonian, one may achieve quantum states that would otherwise be inaccessible.  Moreover, a judicious choice of final Hamiltonian whose groundstate encodes the solution to a problem allows adiabatic transport to be used for universal quantum computation. However, the dephasing effects of the environment limit the quantum correlations that an open system can support and degrade the power of such adiabatic computation. We quantify this effect by allowing the system to evolve over a restricted set of quantum states, providing a link between physically inspired classical optimisation algorithms and quantum adiabatic optimisation. This new perspective allows us to develop benchmarks to bound the quantum correlations harnessed by an adiabatic computation. We apply these to the D-Wave Vesuvius machine with revealing - though inconclusive - results.
\end{abstract}

\maketitle

Nature does remarkably well at performing complex optimisations. These occur on all scales and in all sciences; from evolutionary processes optimising a species for a particular ecological niche, to atoms combining to find low energy crystalline structures. On the other hand, many of the most challenging problems of computation involve performing complex optimisation and algorithms inspired by natural processes have been very successful in solving them\cite{Holland1975,Hopfield1982,Kirkpatrick1983}. 
Amongst algorithms inspired by processes occurring in the physical sciences, many can be summarised in the Langevin equation for the Brownian motion of a particle:
\begin{eqnarray}
&&
\ddot {\bf x} +\gamma \dot {\bf x} + \partial_{\bf x} V({\bf x}) = {\boldsymbol \eta},
\nonumber\\
&&
\langle \eta_i(t)\eta_j(t') \rangle = 2 \gamma T \delta_{ij} \delta(t-t').
\label{LangevinEquation}
\end{eqnarray}
This equation describes the motion of a particle at position ${\bf x}$ in some potential $V({\bf x})$ under the dissipative (described by $\gamma \dot {\bf x}$) and fluctuation (described by the random force, ${\bf \eta}(t)$) effects of a thermal environment. A complex optimisation of many degrees of freedom may be represented by many dimensions of the vector ${\bf x}$.
Perhaps foremost amongst these physics-inspired algorithms is {\it simulated annealing}\cite{Kirkpatrick1983}. This mimics the remarkable fidelity with which a collection of atoms may form complicated crystals by simply heating them up and then cooling down slowly. The simulated annealing algorithm essentially involves integrating  equation (\ref{LangevinEquation}) forwards in time whilst slowly reducing the amplitude of the random forcing. Setting the fluctuating force to zero reduces the Langevin dynamics to a {\it gradient descent}. 

However, it is the possibility of {\it adiabatic computation} in which we are interested here\cite{Kadowaki1998,Brooke1999,Farhi2001,Santoro2002,hastings2009quantum,HastingsFreedman2013}. Put simply, this algorithm involves slowly changing the potential that the system experiences, from a starting potential in which it readily settles to the minimum energy state, to a target whose optimum configuration we require. Classical examples of adiabatic computation are of limited usefulness. As an illustrative example, 
one might think of a toy in which a ball bearing must be guided around a maze by gentle tilting. This is an example of a supervised adiabatic algorithm. An unsupervised version would amount to using a predetermined sequence of tilts and would not be a very successful way of solving the maze. However, it is a remarkable feature of quantum mechanics that if a system is isolated from its environment, the quantum adiabatic theorem implies that such a protocol will always work - though it must be carried out very slowly for the hardest problems\cite{Farhi2001}. 
Here, we investigate how the power of adiabatic optimisation is enhanced by using limited quantum resources. We do so by 
including entanglement into an analogue of equation (\ref{LangevinEquation}). The resulting picture of open-system quantum dynamics enables an appealing geometrical interpretation of the limitations of adiabatic computation; the existence or otherwise of a connected adiabatic computational path. It also suggests a useful alternative perspective upon 
computational classes. Moreover, we use it to develop benchmarks that can quantify the degree of quantum mechanics employed by a putative adiabatic quantum computation. 

Using a given system to perform adiabatic optimisation may involve negotiating  the combined effects of dissipation, noise and intrinsic dynamics. 
Indeed, depending upon the balance between them, the same computational path may be followed in different ways.
A purely adiabatic calculation balances the intrinsic dynamics against a slowly varying potential. Dissipation renders this slow dynamics over-damped, balancing the varying potential against viscous timescales. Turning on fluctuations causes diffusion along the vicinity of the adiabatic path. 
This can make the task of determining the degree of quantum mechanics embodied in a system difficult. In the final section, we apply our ideas to computation using the D-Wave Vesuvius machine. Although spectroscopic probes have shown evidence of entanglement\cite{lanting2014entanglement}, our tests proved inconclusive, because of the relatively high temperature of operation.

\section{Results}
In order to provide a concrete focus to our discussion, 
we work with the D-Wave protocol optimising an Ising Hamiltonian, ${\cal H}_{\hbox{Target}}$ by slowly tuning to it from an initial transverse field Hamiltonian, ${\cal H}_{\hbox{Start}}$:
\begin{eqnarray}
{\cal H}_{\hbox{Target}}&=& \sum_{\langle ij \rangle} J_{ij} \sigma^z_i \sigma^z_j + \sum_i h_i \sigma^z_i
\nonumber\\
{\cal H}_{\hbox{Start}}&=&  \sum_i \Delta \sigma^x_i.
\label{TheProblem}
\end{eqnarray}
Our aim is to develop a Langevin description of the dynamics of this system, into which we can systematically introduce the effects of quantum mechanical entanglement. For analytical and numerical convenience, we focus upon one-dimensional problems. In this case, matrix product states provide both a way to faithfully model the quantum dynamics of a system with finite entanglement resource and a way to quantify that resource through the Schmidt rank. First, we must develop a description valid in the absence of entanglement.

\subsection{Langevin Dynamics of Qubits}
Assuming that coupling to the environment is sufficiently weak that local superposition of up and down states may be supported, we may characterise the state of the system by an $O(3)$ unit vector 
${\bf n}=(\cos \phi \sin \theta, \sin \phi \sin \theta, \cos \theta)$
 for each spin. This unit vector parameterises a coherent superposition of the up and down states by 
 $
|{\bf n} \rangle = \cos  \theta/2 |\uparrow \rangle +e^{i \phi}\sin \theta/2  |\downarrow \rangle.
$
A Langevin equation describing the evolution of such spin coherent states in the presence of a local coupling to the environment is the well-known Landau-Lifshitz-Gilbert equation\cite{Landau1935,Gilbert1955}:
\begin{eqnarray}
&
{\bf n}_i \times \left[ \dot {\bf n}_i+ \gamma {\bf B}_i \right]
= {\bf B}_i + \boldsymbol \eta(t)
&
\nonumber\\
&
{\bf B}_i = \partial {\cal H}( \{ {\bf n}_i \} )/\partial {\bf n}_i ,
&
\label{LLG}
\end{eqnarray}
where the intrinsic, precessional dynamics are described by the first terms on the left and right. The second term on the left describes the dissipative effect of the bath and relaxation of  ${\bf n}_i$ towards the direction parallel to the effective field. The final term on the right describes a random field due to fluctuations in the environment. These fluctuations are independent on each site and are related to the dissipation by 
$\langle \eta_\alpha (t)\eta_\beta (t') \rangle = 2 \gamma T \delta_{\alpha \beta} \delta(t-t')$.

Rather than simply a product state over spin coherent states, equation(\ref{LLG}) represents a mixed state with residual quantum correlations characterised by the product state. This is evident in its derivation, for example from a Keldysh path integral over spin coherent states\cite{Kamenev2011}. This interpretation will be informative when we introduce entanglement, in the next section.

The strong dissipation limit of equation(\ref{LLG}) recovers the model of Ref.\cite{Smolin2013} for the D-Wave machine. In this limit, dissipation rapidly causes a gradient descent to the local minimum of the Hamiltonians. The same adiabatic path is followed through the computation although the strong dissipation suppresses the possibility of non-adiabatic effects. Moreover, introducing thermal fluctuations renders the dynamics diffusive over the computational path. When coarse-grained in time, this recovers the Metropolis dynamics of Ref.\cite{Shin2014}, although over $O(3)$ rather than $O(2)$ vectors.

\subsection{Introducing Entanglement}
In order to faithfully and systematically describe the dynamics of an open system that retains a finite degree of entanglement, we require a suitable class of variational wavefunctions. Matrix product states and their higher dimensional analogues provide such a class\cite{Perez2006}. They can be understood explicitly as a construction based upon Schmidt decomposition at each bond\cite{Vidal2003}. We will use the maximum Schmidt rank as a convenient quantification of the degree of quantumness in a putative adiabatic quantum computer. Whilst there are other ways to quantify the quantumness and potentially other variational wave functions that embody them, matrix product states have some advantage due to the efficient, established ways of simulating their dynamics. 

A matrix product state extends the notion of a product state, 
\begin{equation}
|\psi_A \rangle = \sum_{\{ \sigma \} } A_1^{\sigma_1} A_2^{\sigma_2} A_3^{\sigma_3} ... |\sigma_1,\sigma_2,\sigma_3... \rangle,
\label{MPS}
\end{equation}
by allowing the coefficients $\{ A \}$ to carry auxiliary tensor indices that are contracted between sites. The rank of these tensors corresponds to the Schmidt rank on each bond in the chain\cite{Vidal2003}. equation(\ref{MPS}) describes states formed by a sum over a discrete set of product states corresponding to the different values of the contracted indices. Since any state in Hilbert space can be represented as a sum over product states, in the limit of the Schmidt rank becoming sufficiently large\footnote{For a finite size system, there is a maximum rank at which the matrix product states cover the whole of Hilbert space. Indeed, this maximum rank increases from the ends of the chain and the tensors $\{ A \}$ are in general rectangular.}, an arbitrary state in Hilbert space may be represented. In general, however, the matrix product states of a particular Schmidt rank form a restricted sub-manifold of states in the full Hilbert space\footnote{Strictly, they form a fibre bundle due to their parametrisation invarience\cite{Haegeman2012}}. This notion 
is  sketched in Fig.1.

\begin{figure}
\includegraphics[width=3in]{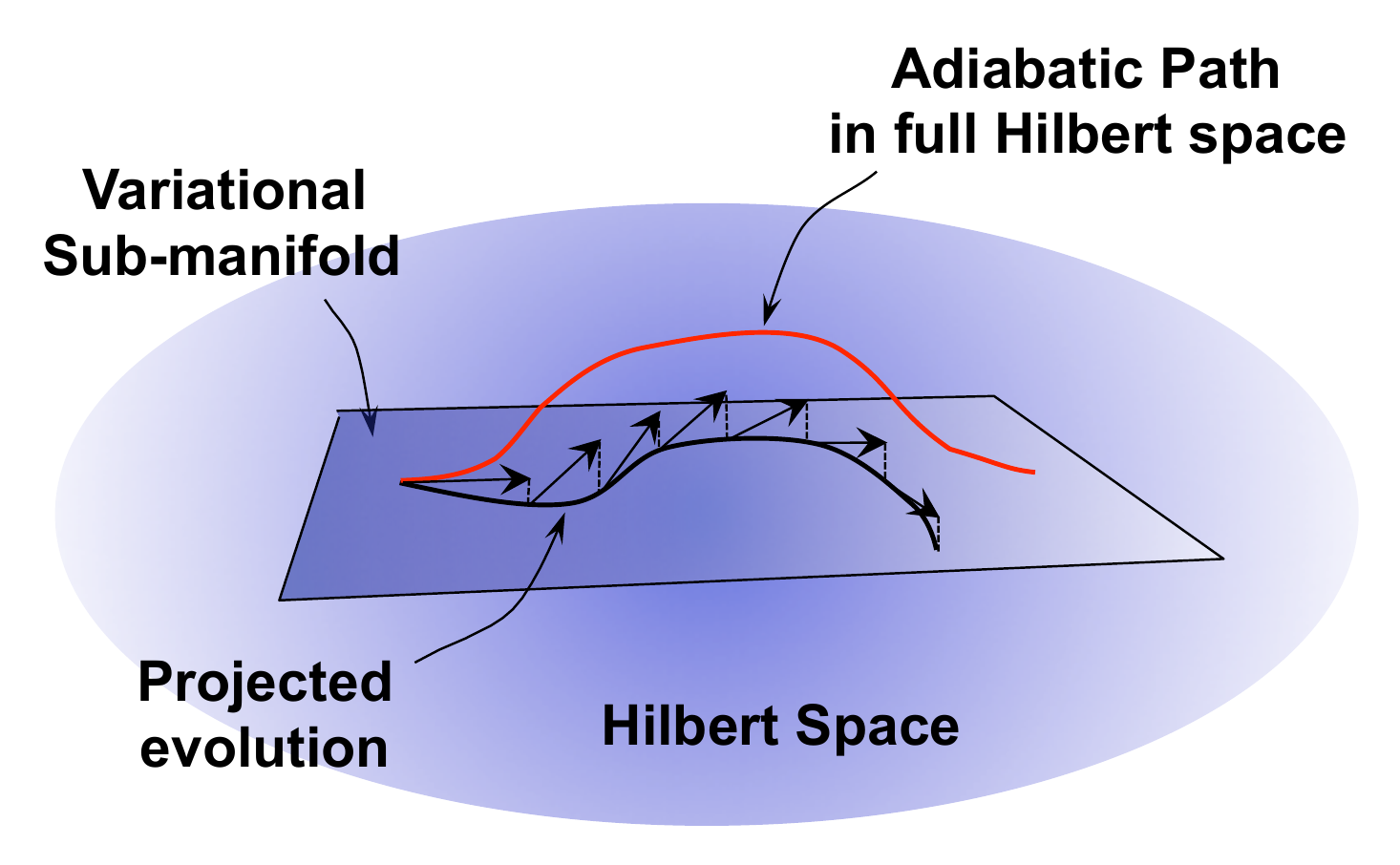}
\caption{\label{fig:VariationalManifold} {\it Variational Sub-Manifold:} In Eqs.(\ref{LLG}) and (\ref{MPSLangevin}), we consider dynamics that are continuously projected onto a variational sub-manifold of a certain degree of entanglement, captured by matrix product states of a particular Schmidt rank. Hamiltonian evolution will generally increase the degree of entanglement and coupling to a bath will counter this. The result is that the systems dynamics may be described semi-classically on an appropriate variational sub-manifold. When the degree of quantum entanglement captured by the variational manifold is insufficient, the projected evolution may deviate from the actual adiabatic evolution in the full Hilbert space. }
\end{figure}

We can now describe the dynamics of a one-dimensional open system that retains a degree of entanglement. The rationale is that coupling to an external bath leads to decoherence that tends to reduce the bond order required to  model the quantum correlations of the system. This effect is balanced by the natural tendency of evolution under a Hamiltonian to cause the entanglement to grow. A suitable Langevin equation over matrix product states takes the form\footnote{This is the Landau-Lifshitz-Gilbert form. With suitable re-scaling of the fluctuation and dissipation terms it may be written in the Landau-Lifshitz form as
$i \langle \partial_{A' } \psi| \partial_A \psi \rangle
\left( 
 \dot A-
\tilde \gamma_{A A''}
\langle \partial_{A'' }\psi| \hat {\cal H} | \psi \rangle
+i \tilde \eta_A
\right)
=
\langle \partial_{A' }\psi| \hat {\cal H} | \psi \rangle
$.}
\begin{eqnarray}
& &
i  \langle \partial_{A' } \psi| \partial_A \psi \rangle
 \dot A
-\gamma_{A' A} \dot A
=
\langle \partial_{A' }\psi| \hat {\cal H} | \psi \rangle
+\eta_A
\label{MPSLangevin}
\end{eqnarray}
The first terms on the left- and right-hand sides of this equation constitute the time dependent variational principle over fixed bond order matrix product states\cite{Dirac1930,Haegeman2011}. 
For product states, these terms correspond to the free precession described by equation(\ref{LLG}) in the absence of coupling to a bath. The additional terms on the left- and right-hand side describe dissipation and noise, respectively. 

As in the case of the Landau-Lifshitz-Gilbert equation, equation(\ref{MPSLangevin}) should be interpreted as an evolution over a mixed state whose micro-states posses residual quantum correlations that can be faithfully represented by a matrix product state\footnote{A derivation of this {\it via} a Keldysh path integral first requires construction of a field integral over matrix product states [C. A. Hooley, J. Keeling, S. Simon and A. G. Green in progress]}. Going to higher bond order increases the amount of entanglement that can be described and the fraction of Hilbert space that is covered by the variational sub-manifold. 

\subsection{The Geometry of Adiabatic Computation}
\noindent
{\it Adiabatic Connectivity of Variational Manifold:}
With certain caveats, the quantum adiabatic theorem guarantees that there exists an adiabatic path between two points in Hilbert space. In principle then, one may initiate a system in the readily-achievable groundstate of a simple Hamiltonian and change the Hamiltonian continuously to a target whose groundstate encodes a problem of interest. 
In order to maintain adiabaticity, this change must be carried out at a rate slower than the inverse gap, the scaling of which determines whether  a particular computation can be performed efficiently\cite{Santoro2002}. These notions lead to important constraints upon adiabatic computation. However, there is another important way in which such computation might fail in an open system.

In an open system,  dephasing renders only a subset of states in the Hilbert space accessible. Because of this only for certain target Hamiltonians is it possible to achieve the states at every point along the adiabatic path. Alternatively, one might say that  depending upon the target Hamiltonian, the manifold of accessible states may or may not contain the adiabatic path. Both adiabaticity and connectivity of the adiabatic path must be satisfied in order for an adiabatic computation to proceed successfully. It is in pursuing the latter that one might hope to address the thorny  and contentious issue of how quantum mechanical is a given adiabatic computation.

\begin{figure}
\includegraphics[width=3.5in]{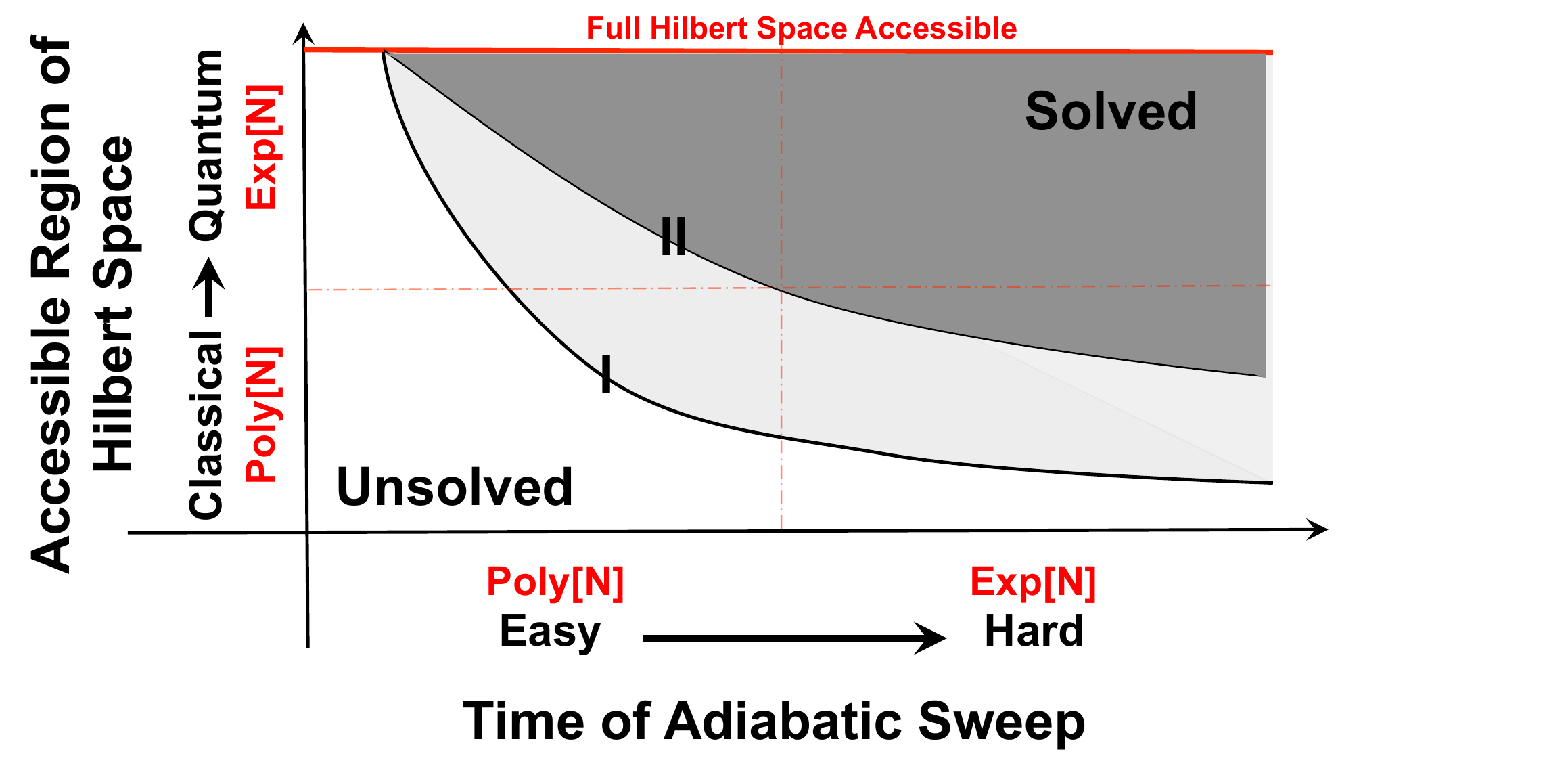}
\caption{\label{fig:SuccessHull} 
{\it Conditions for Successful Adiabatic Computation:}
The boundary between successful and unsuccessful adiabatic computation forms a convex hull in a plot of accessible fraction of Hilbert space {\it versus} sweep time. The minimum computation time, $T^*$, occurs when all of the Hilbert space is accessible. The minimum fraction of Hilbert space for adiabatic computation is given by the horizontal asymptote at $f^*$. Schmidt rank is used as a discontinuous measure of the size of this accessible region in the present work. Contact with measures of computational complexity is made by dividing both axes into regions polynomial and exponential in the system size, $N$. A quantum and classically easy calculation (I) can be performed in polynomial time with only a small fraction of the Hilbert space.
A quantumly easy, classically hard computation (II) can only be performed in a polynomial time using a large fraction of the Hilbert space. 
}
\end{figure}

The combination of these effects can usefully be summarised in a graph of sweep time {\it versus} accessible fraction of Hilbert space as illustrated in Fig.\ref{fig:SuccessHull}. For a closed system, when the whole of the Hilbert space is accessible, the adiabatic computation may be performed\footnote{With exponentially small probability of error.} in a minimum time $T^*$. As the sweep time increases and the fraction of accessible Hilbert space goes down, we expect the line delimiting success from failure to form a convex hull as depicted in Fig.\ref{fig:SuccessHull}. Eventually, a threshold is reached with a minimum fraction, $f^*$, of the Hilbert space accessible so that the problem is just adiabatically soluble. Here, we use Schmidt rank as a discontinuous measure of the size of the accessible region of Hilbert space.

\vspace{0.1in}
\noindent
{\it Quantum and Classical in Adiabatic Computation:}
The model of adiabatic computation in open systems presented in equation(\ref{MPSLangevin}) reduces it to a classical analogue algorithm. Of course, the classical data (the tensors of the matrix product states) contain important information about key quantum correlations, but nevertheless the evolution is classical\cite{Vidal2003}. However, adiabatic and gate based quantum computation are known to be computationally equivalent\cite{Siu2005,Kempe2006,Mizel2007,aharonov2008adiabatic}. How do we reconcile these two positions? The answer is contained in the division of the graph of Hilbert space fraction {\it versus} sweep time into regions polynomial and exponential in system size as in Fig. \ref{fig:SuccessHull}. This division corresponds to easy and hard in sweep time, and to classical and quantum in Hilbert space fraction. 
Problems can be classified according to the quadrants of this graph that the success hull passes through. For example, a quantum and classically easy calculation can be performed in polynomial time with only a small fraction of the Hilbert space. Its success hull passes through the quadrants (a), (c) and (d). A quantum easy, classically hard computation will pass through quadrants (a) and (d). Moreover, a commutation that can only be performed quantum mechanically might have a success hull that passes through quadrants (a) and (b). 

Thus, we see that even though an adiabatic computer may harness significant quantum resources, unless the accessible region of Hilbert space grows sufficiently rapidly with the size of system, the class of soluble problems will be classical. A saturation in the length scale of entanglement in an open system will inevitably result in classical scaling. This appears to be the source of much of the polarisation of the debate about the degree of quantumness in a putative adiabatic computer. Such a computer may harness a significant degree of quantum mechanics in its operation, but  nevertheless be strictly a classical computer. Nevertheless, taking a pragmatic view, the scaling though linear (or polynomial) may have a very large pre-factor that is exponential in the length scale over which the system maintains quantum mechanical coherence. 

\vspace{0.1in}
\noindent
{\it Quantifying Quantumness and the Breakdown of Adiabatic Computability:}
The threshold of adiabatic computability is particularly revealing. This corresponds to the point where, upon reducing the size of the accessible region of Hilbert space, regions of the variational manifold become adiabatically disconnected. In the case where this happens continuously, it corresponds to the point where the excitation gaps close for evolution projected  onto the variational manifold. Viewed as classical dynamics, this may coincide with the adiabatic evolution becoming Lyapunov unstable suggesting an interesting link between the limits of classical and adiabatic computability. The relationship to threshold theorems for gate base quantum computation is an interesting avenue of further study. 

This simple, geometrical idea also contains the seed of how to quantify the degree of entanglement in a notional adiabatic computer. In order to be specific, let us focus upon one-dimensional problems. We classify sets of target Hamiltonians according to the minimum Schmidt rank or bond order of matrix product states\footnote{Schmidt rank gives a fairly crude, discrete measure of the fraction of the accessible fraction of Hilbert space. A more complete, continuous measure might perhaps be constructed from the Schmidt spectrum
} whose zero-temperature, projected adiabatic evolution  gives the correct solution - we could even refer to this as the rank of that particular problem. Testing our adiabatic computer with problems of different rank we expect to find a step change in computability corresponding to the degree of entanglement that is supported by the notional adiabatic computer. In the following sections, we identify a set of such one-dimensional test problems, characterise them in this way and report upon their use as benchmarks for the D-Wave Vesuvius computer.

\subsection{Developing a Set of Test Problems}

We now use these ideas to develop a set of test problems. These will comprise  of a set of one-dimensional Ising Hamiltonians of the type given in equation(\ref{TheProblem}), grouped according to the minimum Schmidt rank, $\chi^*$,  required to solve them adiabatically starting from a transverse field Hamiltonian. For each target Hamiltonian, we i. determine its minimum energy state by an exhaustive search, ii. follow its projected evolution in a variational manifold parametrized by matrix product states of fixed Schmidt rank using equation(\ref{MPSLangevin}). Since our aim is to characterise the zero-temperature adiabatic computability of the target Hamiltonian, this is carried out in the absence of noise. We use a time-evolved block decimation algorithm for our explicit computation - such evolution agrees with that obtained from the time-dependent variational principle in the limit of short time steps\cite{Haegeman2011}. iii. identify the minimum bond order, $\chi^*$, at which the projected evolution takes the system to the exact minimum. 
We emphasise that time-evolving in this way is not an attempt to simulate any particular system. Rather, it provides a bound upon the quantum resources required to minimise a given target Hamiltonian adiabatically. 

Our choice of target Hamiltonian is restricted in several ways. We must be able to find its ground state exactly in reasonable time by exhaustive search. It must also be embeddable on the chimera graph used by D-Wave machines.  We have considered 500 randomly chosen Hamiltonians on spin chains of length 100; the couplings are chosen randomly from the set $J \in \pm \{ 0.2,0.4,0.6,0.8,1 \}$ and the local fields from the set $h \in \pm \{0, 0.2,0.4,0.6,0.8,1 \}$. Although the ground states of these target Hamiltonians are relatively easy to find, they provide a surprisingly tough test for adiabatic computation. In the supplemental materials, we outline a similar analysis of the two-leg ladder, which proved considerably less discriminating. 

Surprisingly, although groundstates of the transverse field Ising model may require high bond orders for a faithful representation using matrix product states, we find that the threshold for adiabatic computability is at Schmidt rank 1 or 2 in all cases\footnote{Except for a handful of problems with which our code experienced numerical difficulties}. In order to place each problem instance on a graph following Fig.\ref{fig:SuccessHull}, we ran our simulations at Schmidt ranks $1,\;2,\;3$ and $4$ and on a coarse grid of times $T_0$, $2T_0$, $4T_0$, $8T_0$ (where $T_0$ was an arbitrary large multiple of the fundamental intrinsic timescale of the Josephson junction).

\subsection{Entanglement in the D-wave Vesuvius Machine}
The D-wave Vesuvius system is a large-scale Josephson junction array with tuneable and controllable couplings. It realises  512 quantum bits on a chimera graph and by controlling their interactions can follow the adiabatic protocol described above\cite{Johnson2011}. Spectroscopic probes have demonstrated that it supports entanglement\cite{Berkley:2013uq,lanting2014entanglement} within its fundamental 8-bit cluster. However, evidence of this entanglement in its performance as a computer has proven elusive\cite{Johnson2011,Boixo2013,Smolin2013,Wang2013,Boixo2014evidence,Shin2014}. Such efforts have typically focussed upon comparing entirely classical or entirely quantum models ({\it i.e.} with every point in the Hilbert space accessible) with the performance of Vesuvius. Perhaps unsurprisingly the resulting debate has been rather polarised.  Our approach potentially provides a route to black-box benchmarking that lies between the quantum and classical extremes.

\begin{figure}
\includegraphics[width=3in]{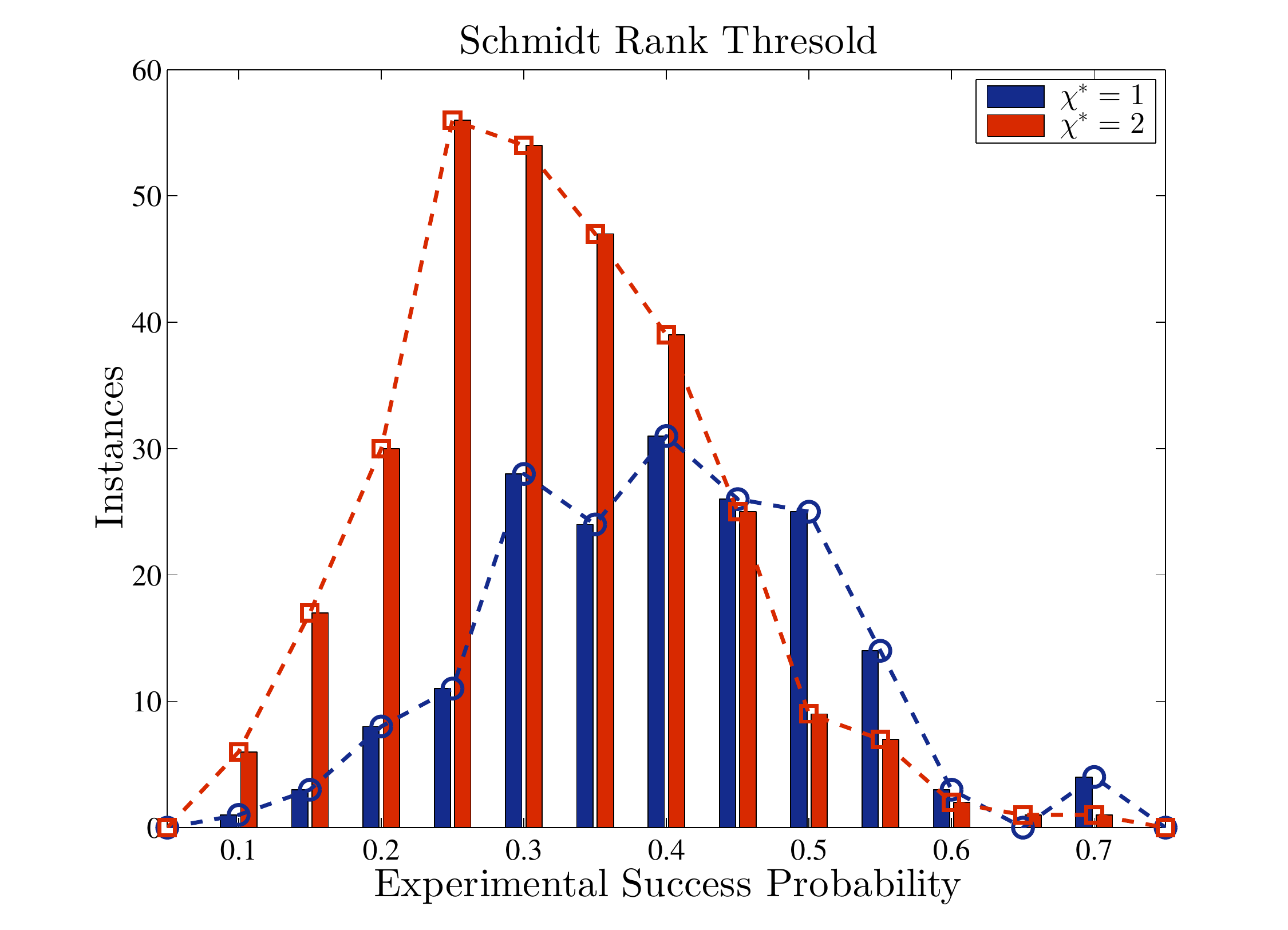}
\caption{{\it   Success probability correlated with minimum Schmidt rank required for successful adiabatic computation:} Histograms of number of problem instances versus D-Wave success probability. The data are divided into cohorts requiring different  Schmidt rank for successful adiabatic computation.
We show data for 500  instances of the one-dimensional chain.  Problem instances with  higher threshold Schmidt rank - and so more quantum according to Fig.\ref{fig:SuccessHull} - are more difficult to solve on the D-wave Vesuvius machine. }
\label{fig:Results1}
\end{figure}

\begin{figure}
\includegraphics[width=3in]{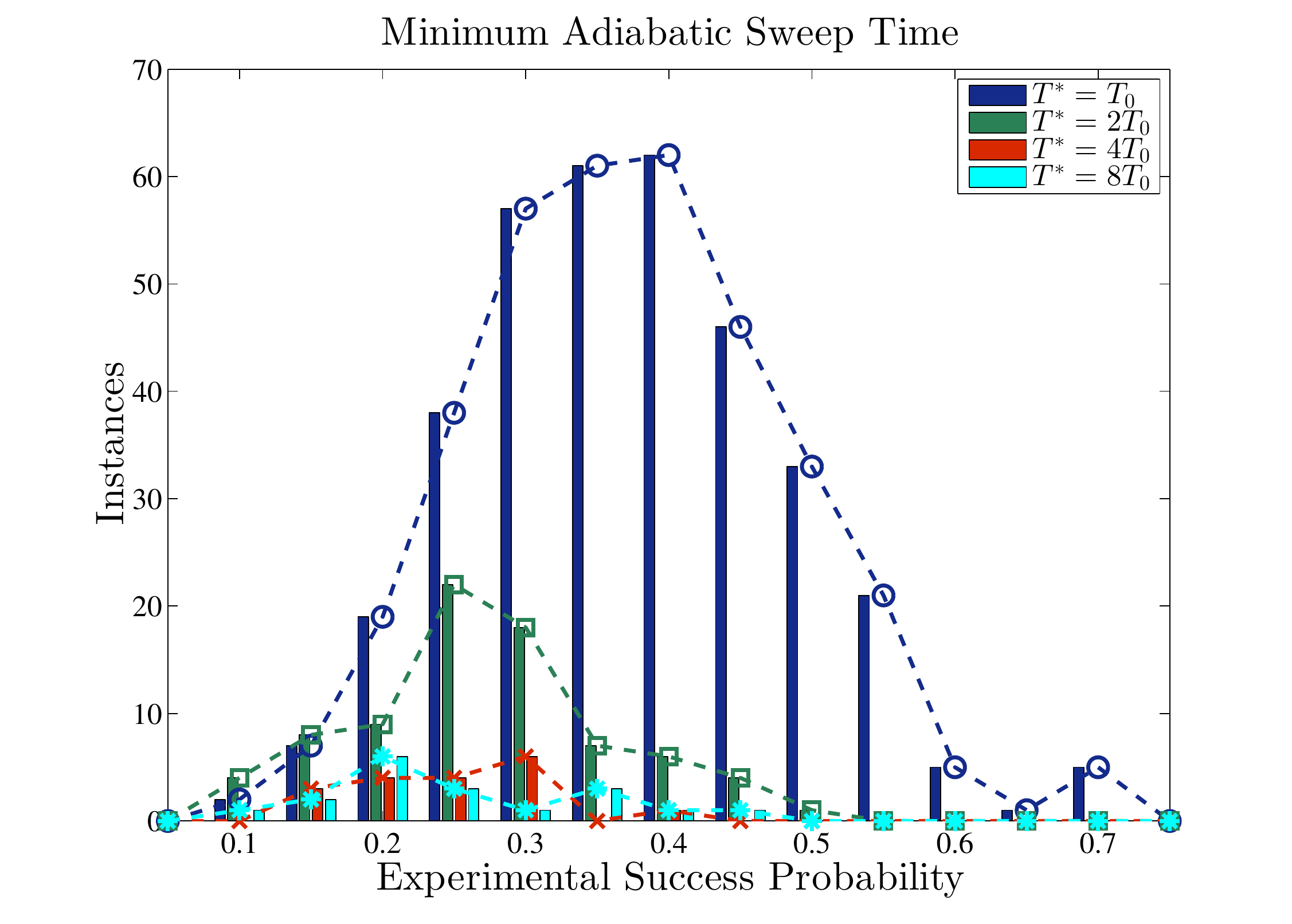}
\caption{{\it   Success probability compared with minimum sweep time required for successful adiabatic computation:} 
Histograms of number of problem instances versus D-Wave Vesuvius success probability. The data are divided into cohorts requiring different sweep rates for successful adiabatic computation at Schmidt rank 2.
We show data for 500  instances of the one-dimensional chain. Problem instances requiring longer sweeps - and so harder according to Fig.\ref{fig:SuccessHull}  - are more difficult to solve on the D-wave Vesuvius machine.   }
\label{fig:Results2}
\end{figure}

We have tried our test problems on the D-Wave Vesuvius machine at USC, running multiple attempts and embeddings of each problem instance in order to determine the probability of its successful computation. Our results are summarised in Figs. \ref{fig:Results1} and \ref{fig:Results2}. Fig. \ref{fig:Results1} plots a histogram of number of problem instances versus success probability separating the problem instances according to their threshold Schmidt rank, $\chi^*$, for adiabatic computability. Fig. \ref{fig:Results2} also plots a histogram of number of problem instances versus success probability, this time separating problem instances according to their minimum adiabatic sweep time ( $T_0$, $2T_0$, $4T_0$, $8T_0$) at bond order 2. 
The probability of successful computation on Vesuvius shows only a weak correlation with the entanglement required for adiabatic computability and a similarly weak correlation with the difficulty of the problem determined by the minimum theoretical sweep rate. The cause of this appears to thermal fluctuations\footnote{Technical issues concerning the accuracy with which the fields and couplings can be tuned do not affect our results. This can be quantified  straightforwardly by plotting auto-correlations of the system's performance for different embeddings of the target Hamiltonian [see supplemental materials] and may be mitigated by gauge averaging\cite{Boixo2014evidence}. We have checked our calculations with and without such gauge averaging and find no change to our results. }.
It is very difficult to disentangle thermal and quantum effects and this ultimately renders our results inconclusive.

At the lowest temperatures, we expect a bimodal division of success or failure depending upon whether the degree of entanglement has passed threshold. In this limit, increasing the temperature will be detrimental as thermal fluctuations cause deviations from the adiabatic path. However, at higher temperatures, even if it does not support sufficient entanglement, an adiabatic computer can solve a  problem by thermal diffusion between disconnected regions of the accessible manifold\cite{amin2008thermally}. This is consistent with the weak correlation of D-Wave success probability with minimum sweep time shown in Fig.\ref{fig:Results2}. 

Ideally, we would like to reduce thermal fluctuations by lowering temperature. However, the D-Wave Vesuvius operates at the limits of refrigeration technology and this was not possible. Runs using the trick of lowering the energy scales to mimic higher temperature\cite{Vinci2014distinguishing} did not yield clear results. Changing the D-Wave sweep time permits an alternative by changing the time for which the system is exposed to thermal fluctuations. Increasing sweep times resulted in all success probabilities going down or remaining unchanged. This provides circumstantial evidence that the computation does indeed proceed using entanglement resources and that thermal fluctuations are detrimental to this. Unfortunately, we were not able to shorten the sweep time in order to confirm this.

\section{Discussion}

We have developed a new perspective upon adiabatic computation in open systems.
In the absence of some scheme of quantum error correction, the decohering effects of the environment determine a maximum degree of entanglement that a system can support. Motivated by this, we have focussed upon zero-temperature simulation over variational states with a a quantified degree of entanglement. This leads to a new way of quantifying constraints upon adiabatic computability. 

The {\it adiabatic success hull} delimits the region in a plot of accessible fraction of Hilbert space {\it versus} sweep time where a given problem is adiabatically soluble. We have suggested how it can be related to computational complexity classes. Viewed in this way, the threshold of adiabatic computability may bear relation to both error thresholds for the performance of gate based quantum computation and to Lyapunov instability of semi-classical dynamics. We have identified sets of one-dimensional problems with different threshold Schmidt rank (used here as a proxy for the accessible fraction of Hilbert space) that provide benchmarks for the degree of quantum mechanics in a putative quantum computer. 

The D-Wave Vesuvius is the first large-scale controllable Josephson array with the potential to implement an adiabatic optimisation scheme. Although spectroscopic experiments have demonstrated the existence of entanglement\cite{Berkley:2013uq,lanting2014entanglement} in this system, its signatures have been harder to see in the results of computation\cite{Johnson2011,Boixo2013,Smolin2013,Wang2013,Boixo2014evidence,Shin2014}. 
Applying our benchmarks to Vesuvius machine is inconclusive because of the finite temperature at which it operates. The interplay of thermal and quantum effects can in principle be captured by the quantum Langevin equation over matrix product states that we propose in equation(\ref{MPSLangevin}). Deriving measures that allow one to differentiate their effects as well as study their interplay is an important future goal.

There are several natural directions for future development of this work. We have used the Schmidt rank as a measure of the size of the variational manifold. A continuous measure would be preferable and it may be possible to use the Schmidt spectrum to develop such a measure - fixing the Schmidt rank to a particular value is, after all, a discrete constraint upon the Schmidt spectrum. Alternative variational wave functions may also be explored. Further development of the relationship of the adiabatic success hull to conventional measures of computational complexity, and particularly the relationship of the adiabatic threshold to similar bounds for gate-based quantum computation would be very revealing. As we noted in Section IV, this threshold may also be related to classical computability and the Lyapunov stability of the dynamics on the restricted manifold\cite{Shin2014} described by the Langevin equation. Extending our analysis to higher dimensions requires the development of time-dependent codes for suitable tensor networks. This, of course, presents interesting challenges beyond the study of adiabatic computation. 
From the experimental point of view, it is hoped that future generations of the D-Wave device will permit more general choice of control parameters ({\it e.g.} target hamiltonian, annealing time and temperature). Using the approach advocated here should allow the machine to be tuned to make most efficient use of its quantum resources. 

As a final comment, we reiterate that without quantum error correction, adiabatic computation using an open system will strictly be classical since it may be described by a Langevin equation. This is the case even if a significant degree of entanglement is harnessed. However, utilising entanglement up to a given length scale will generate a Schmidt rank exponentially large in that scale and will very rapidly become practically beneficial.

\textbf{Acknowledgment:}  The authors benefited from stimulating discussions with Nick Chancellor, Gabriel Aeppli, Vadim Oganesyan, Chris Hooley, Jonathan Keeling, Steve Simon and Simone Severini. We have made extensive use of matrix product evolution code written by Felix Nissen. We acknowledge support from the EPSRC through grant numbers EP/H005544/1, EP/K02163X/1, EP/I004831 and the TOPNES programme grant EP/I031014/1 and thank the ISI at the USC Viterbi School of Engineering and Lockheed-Martin for access to the D-Wave Vesuvius machine.

\appendix

\section{Experimental Methods}
\begin{figure}[h] 
\centering
\includegraphics[width=1 \columnwidth]{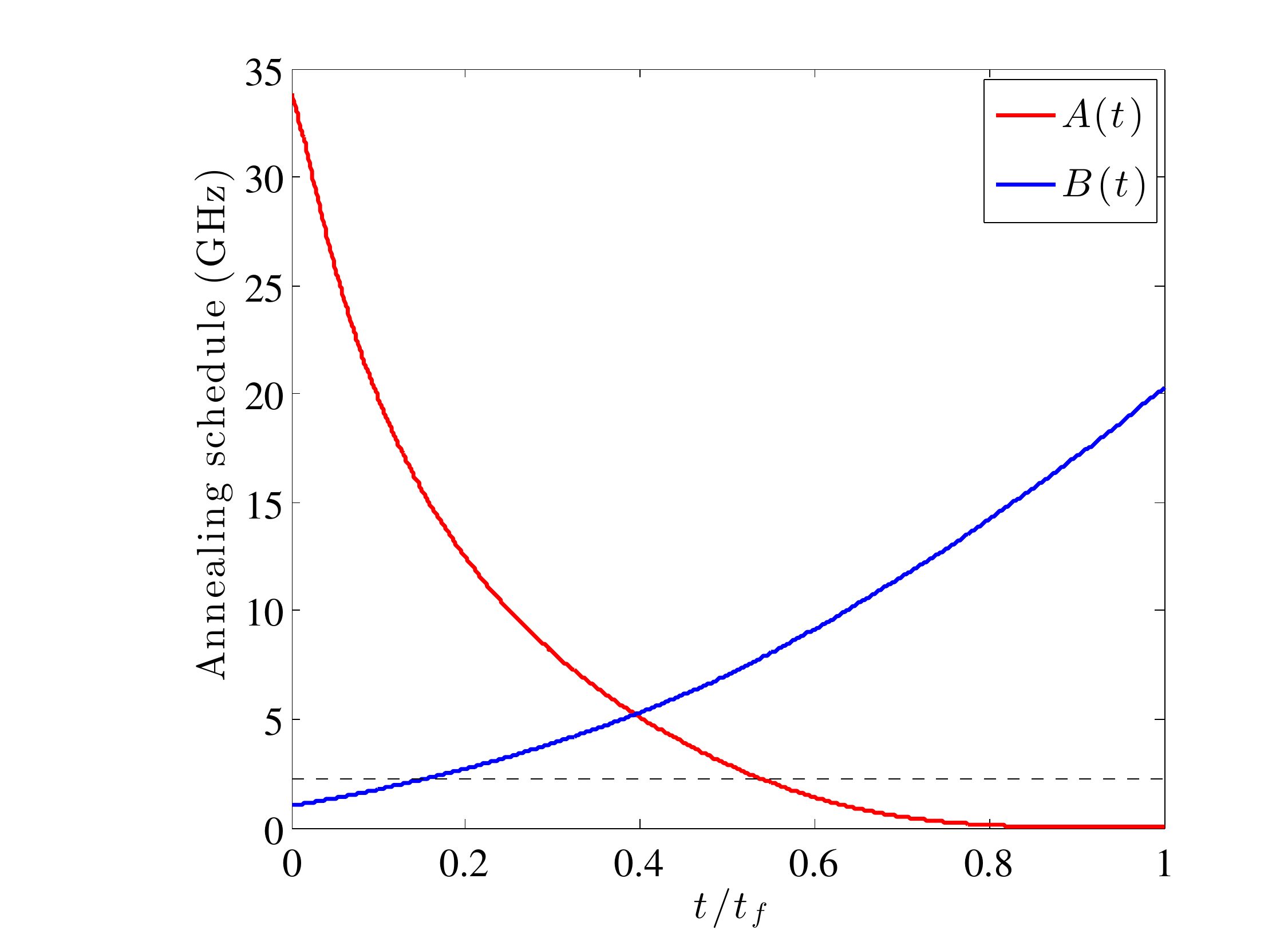}
\caption{{\it Annealing schedule of the D-Wave Vesuvius
device in energy units where} $h=1$:
The black dashed line is the operating temperature ($18$mK) of the device. The large $A(0)/T$ value ensures that the initial state is the ground state of the transverse field Hamiltonian. The large $B(t_f)/T$ value ensures that thermal excitations are suppressed and that the final state reached is stable.}
\label{fig:AnnealingSchedule}
\end{figure}

\begin{figure}[h]
\centering
\includegraphics[width=0.8\columnwidth]{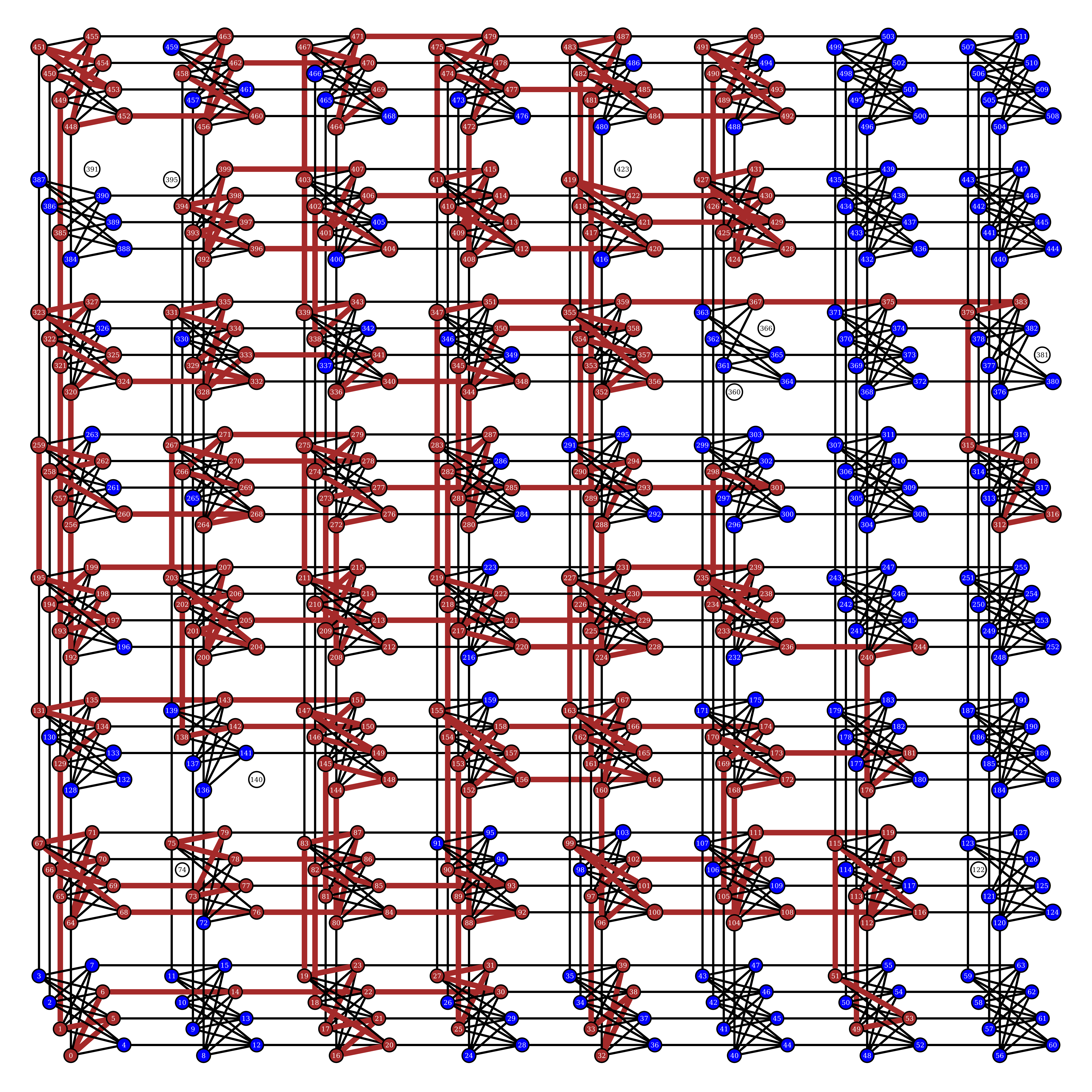}
\caption{{\it Connectivity of the D-Wave Vesuvius device:} 503 flux qubits (in red and blue) are usable. 9 qubits (in white) cannot be reliably calibrated and are not usable. The qubits and the couplers highlighted in red represent 3 different implementations of a 100 spin chain. The embeddings are randomly generated in order to average over the  physical inhomogeneities of the chip. }
\label{fig:Embedding}
\end{figure}

The D-Wave Vesuvius implements the following schedule of quantum annealing: 
\begin{eqnarray}
{\cal H}_{\hbox{Tot}}    =   A(t) {\cal H}_{\hbox{Start}}+ B(t) {\cal H}_{\hbox{Target}}  \, ,
\label{annealing}
\end{eqnarray}
where the driving functions $A(t)$ and $B(t)$ are changed in time  according to the annealing procedure shown in Fig.\ref{fig:AnnealingSchedule}. The target and starting Hamiltonians, $ {\cal H}_{\hbox{Target}}$ and $ {\cal H}_{\hbox{Start}}$,  are defined in equation (2) of the main text. The most general Ising Hamiltonian, $ {\cal H}_{\hbox{Target}}$, that can be implemented in the device is a sub-graph of the device hardware connectivity  - the so-called Chimera  graph shown pictorially in Fig.\ref{fig:Embedding}. Within the limits of the hardware connectivity,  each coupling and longitudinal field can be set to an arbitrary value chosen within the ranges $|J_{ij}|<1$ and $|h_i|<2$ respectively. This allows for the presence of  $\sim1400$ independent tuneable parameters in
the target Hamiltonian.

\begin{figure}[hb]
\centering
\hspace{-0in}\includegraphics[width=0.9\columnwidth]{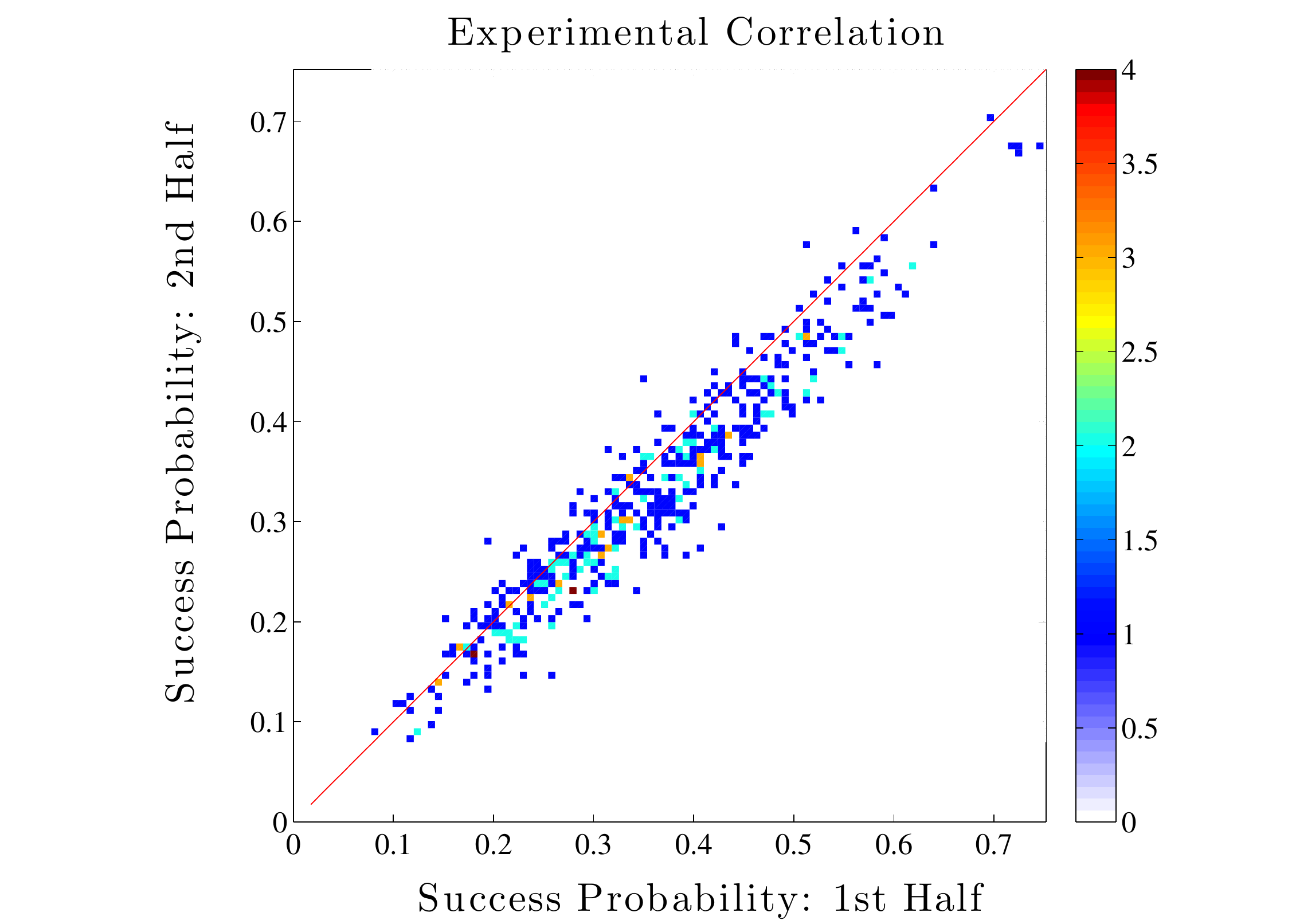}\caption{{\it Autocorrelation of Data:} Success probabilities for two halves of the collected data. The agreement between these is sufficient for the studies that we carry out.}
\label{fig:check}
\end{figure}

{\it Statistical and Systematic Errors:} An important point to consider when carrying out experiments with the D-Wave device is the presence of statistical and systematic errors. Because of these errors, the difference between the actual and intended value of the couplings $J_{ij}$ and $h_i$ can be as large as  $\sim0.05$. To average over statistical errors, we have initialised the device 20 times for each instance.  To average over systematic errors, we have implemented the same instance using different physical spins of the device (different embeddings). For each programming cycle we have taken 1000 readouts from which we have extracted the success probability for all instances. As an example, Fig.\ref{fig:Embedding} shows how the same chain of 100 spins can be embedded in 3 different ways (in parallel and using the available connectivity) on the D-Wave device. We have implemented 
6 different embeddings for the one-dimensional chain and 25 embeddings for the two-leg ladder. In order to check that the statistics of our collected data are sufficiently constrained, we compare the success probabilities for the  chain dividing the data in two cohorts. The results show in Fig.\ref{fig:check}  confirm that experimental errors are under control for the purpose of the present work. All experiments have been performed using the shortest annealing time that can be set on the D-Wave device: $t_f = 20\mu s$.

\section{Annealing time}
\begin{figure}
\centering
\includegraphics[width=\columnwidth]{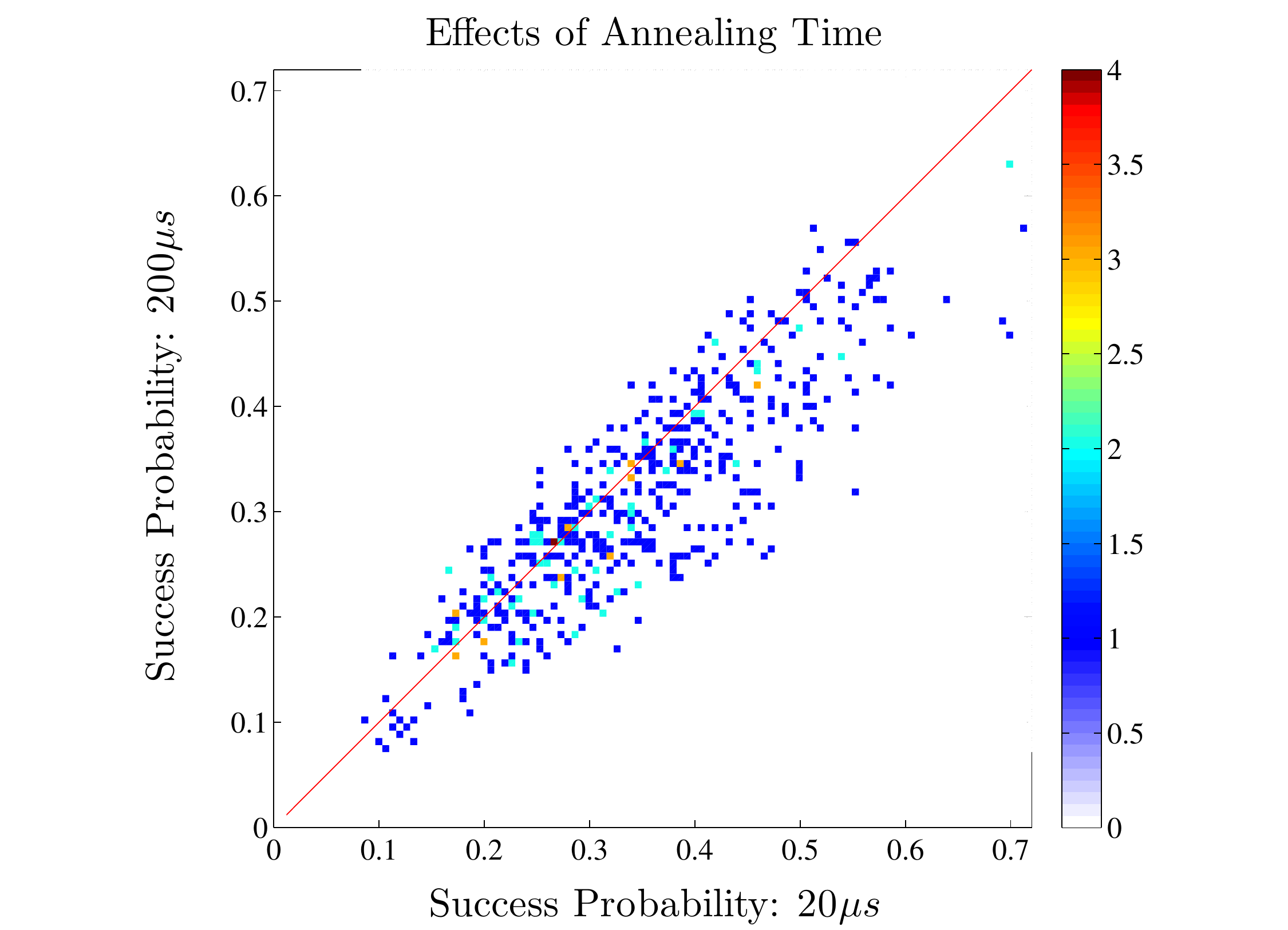}
\caption{{\it Varying Sweep Time:} Comparison between the success probabilities measured at two different values of the total annealing time $T_f$. There is a small reduction in success probability using the longer sweep time.}
\label{fig:annealing}
\end{figure}

As discussed in the main body of the text, it is apparent that computation on the D-Wave device is strongly affected by thermal fluctuations. Although adiabatic computation in a closed system is best carried out over long sweep times in order to avoid non-adiabaticity, in an open system it may be preferable to limit the time of exposure to thermal fluctuations, balancing the errors incurred from non-adiabaticity against a reduction in thermally induced errors. In order to to investigate this balance, it would be desirable  to study how the experimental success probabilities vary 
when the annealing time is changed from a sub-optimal to a super-optimal regime. Unfortunately,  as first noted in Ref.[26] of the main text, 
the minimum adiabatic time accessible to the D-Wave device appears to be super-optimal and cannot be reduced to the sub-optimal regime. We confirmed this observation here
by repeating our experiments on the one-dimensional chains at a larger value of the annealing time: $T_f = 200\mu s$. Fig.~\ref{fig:annealing} shows that the performance of the device reduced slightly with respect to runs with shorter annealing time. 

\section{Ladder Graphs:}
\begin{figure}
\centering
\includegraphics[width=0.8\columnwidth]{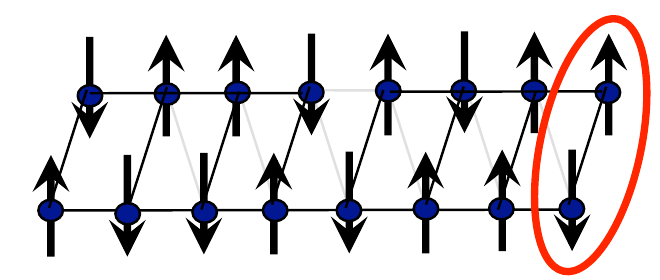}
\caption{{\it Embeddable 2-leg Ladders:} A general two-leg ladder is expected to provide a severe test of adiabatic computability. However, the constraint of embeddibility on the Chimera graphs forbids connections on the links shown in light grey. The remaining links - shown in black - lead to problems that are easily solved by the D-Wave Vesuvius device. The red ellipse indicates the combination of two spins into a 4-level hyper spin over which matrix product states can be constructed}
\label{fig:TwoLegLadder}
\end{figure}

In addition to the one-dimensional chains considered in the main text, we have also studied the adiabatic solubility 
of target Ising Hamiltonians on two-leg ladders as shown in Fig.\ref{fig:TwoLegLadder}. Ladder graphs are expected to provide a more severe test, since the loops that they contain can lead to frustration. Unfortunately, the restrictions of the chimera graph are such that an arbitrary ladder cannot be embedded. Neither can we embed graphs with loops containing odd numbers of spins (those that we expect to be frustrated) nor an arbitrary number of even-spin loops. We therefore consider the simplified graph of 16 spins illustrated in Fig.\ref{fig:TwoLegLadder}. We chose 1000 instances of random couplings and fields on this graph from the sets $J \in \pm \{ 0.2,0.4,0.6,0.8,1 \}$ and  $h \in \pm \{0, 0.2,0.4,0.6,0.8,1 \}$.

\begin{figure}
\vspace{0.25in}
a)
\includegraphics[width=2.75in]{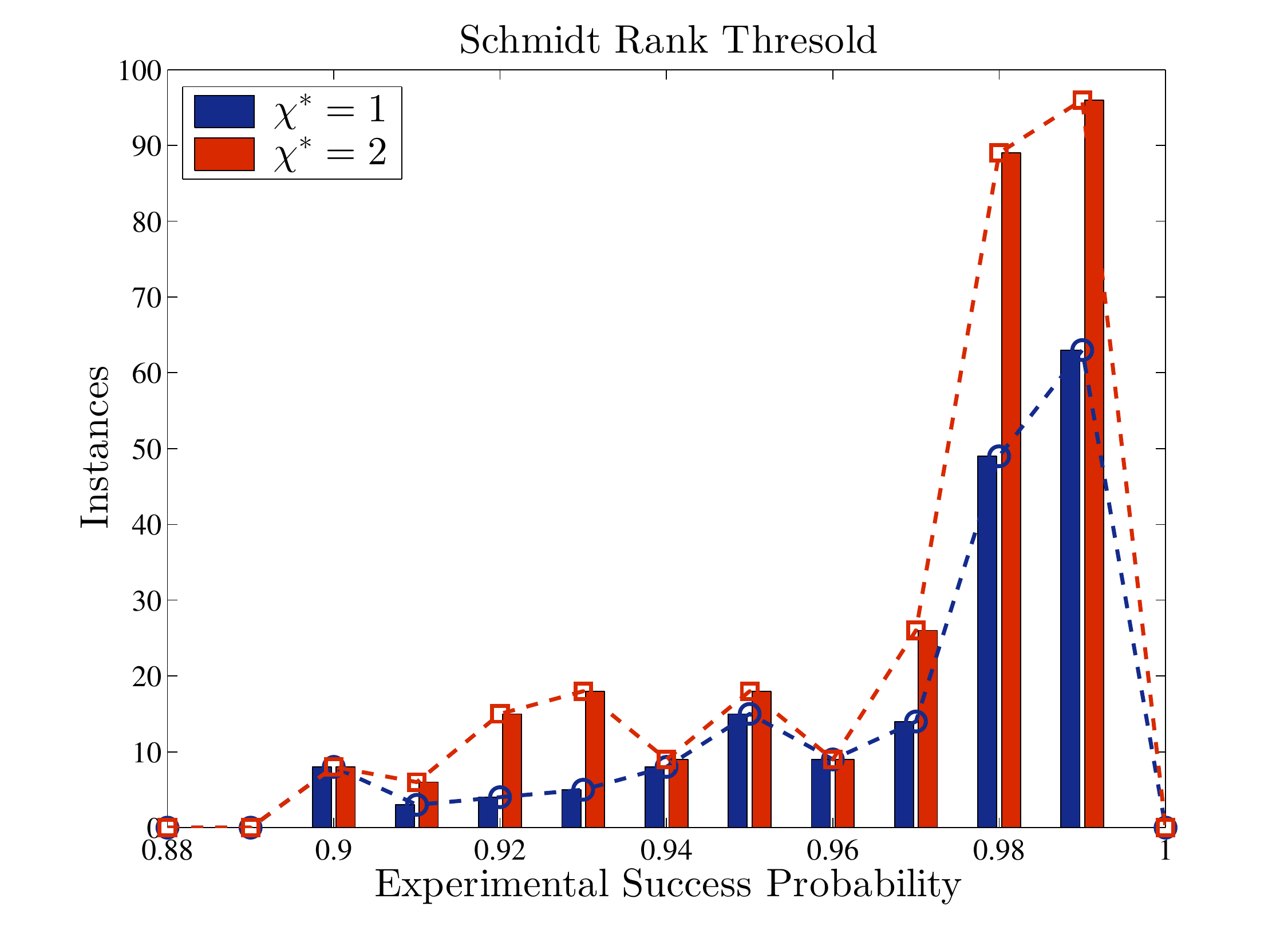}
b)
\includegraphics[width=2.75in]{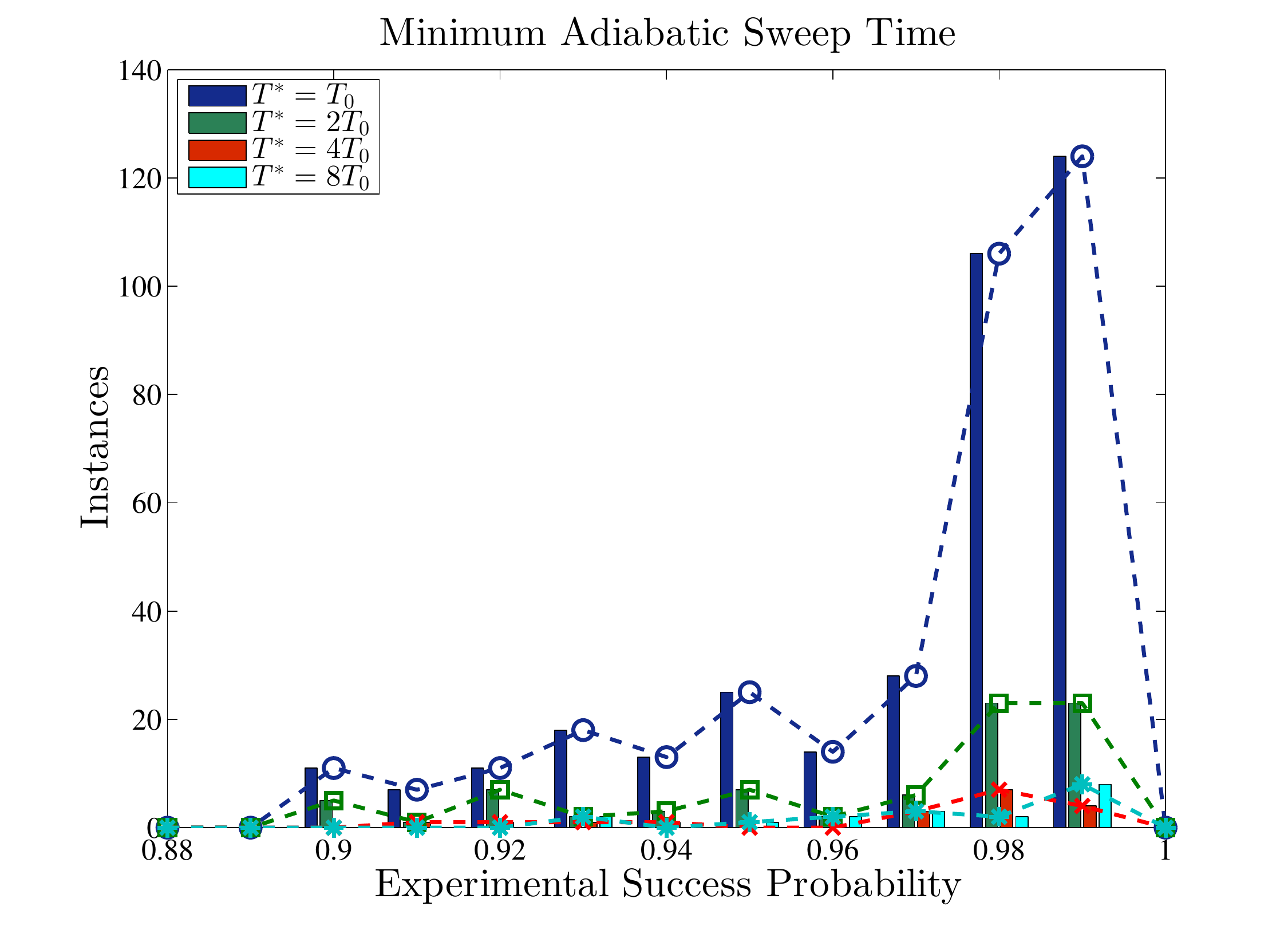}
\caption{{\it   Success probabilities for the Two-leg Ladder:} Histograms of number of problem instances versus D-Wave success probability for instances of the two-leg ladder. In a), the data are divided into cohorts requiring different  Schmidt rank for successful adiabatic computation. In b), the data are divided into cohorts requiring different sweep rates for successful computation at Schmidt rank 2. The two-leg ladder is considerably easier for the D-Wave Vesuvius machine to solve than the one-dimensional chain. There is no discernible correlation between success probability and either Schmidt rank or sweep time. 
}
\label{fig:LadderResults}
\end{figure}

One subtlety in simulating the time-dependence of spins on this graph is that one cannot use a simple matrix product state over spins. Instead, we combine spins on one rung of the ladder to form a single, 4-level hyper spin over which we construct matrix product states. Fig.\ref{fig:LadderResults} shows the result of attempting to solve these problems on Vesuvius, divided into cohorts according to a) threshold Schmidt rank and b) minimum sweep time. These problems are much more easily solved by Vesuvius than the one-dimensional cases. Moreover, performance shows no correlation with Schmidt rank or sweep time.

\bibliography{Article}

\end{document}